\documentclass{ltxdoc}

\usepackage[numbered]{hypdoc}
\usepackage{hologo}
\usepackage{hyperref,xcolor}
\usepackage{longtable,booktabs}
\usepackage{academicons}

\usepackage{verbatim}
\usepackage[english]{babel} 
\usepackage{booktabs}
\usepackage[utf8]{inputenc} 
\usepackage{todonotes}
\usepackage{natbib}
\usepackage[linesnumbered,ruled]{algorithm2e}
\usepackage{enumitem} 
\usepackage{nomencl}
\usepackage{subfigure}
\usepackage[flushleft]{threeparttable}
\usepackage{color, colortbl}
\usepackage{scalefnt}
\usepackage{adjustbox}
\usepackage{framed}
\usepackage{float}
\usepackage{xspace}
\usepackage{flushend}

\newcommand{\totalMembros}{252\xspace}
\newcommand{\numEdicoes}{21\xspace}

\newcommand{\MyBox}[1]{\vspace{3mm}\noindent\framebox[\columnwidth][c]{\parbox[b]{0.95\columnwidth}{ #1 }}}

\definecolor{blue}{rgb}{0.19,0.31,0.54}
\hypersetup{colorlinks=true, linkcolor=blue, urlcolor=blue, hyperindex}

\usepackage{totcount}
\newtotcounter{IconsCounter}
\EnableCrossrefs
\CodelineIndex
\RecordChanges

\changes{v1.8.6-2}{2018/06/27}{Fixed bugs in a few icon commands.}
\changes{v1.8.6-1}{2018/06/25}{Removed \texttt{fontspec} dependency.}
\changes{v1.8.6}{2018/04/03}{Release of v1.8.6.}
\changes{v1.8.3}{2018/04/03}{Release of v1.8.3.}
\changes{v1.8.0-1}{2018/03/27}{Enhancement release: adjusted \texttt{\textbackslash{}newfontfamily} to account for the font installed via \texttt{tlmgr}.}
\changes{v1.8.0}{2017/05/15}{Major release of v1.8.0.}
\changes{v1.7.2}{2016/06/12}{Fix release: corrected a bug in the style file.}
\changes{v1.7.1}{2016/06/12}{Fix release: corrected a bug in the style file.}
\changes{v1.7.0}{2016/05/16}{Major release of v1.7.0.}
\changes{v1.6.0-1}{2016/03/02}{Better documentation.}
\changes{v1.6.0}{2016/01/30}{Major release of v1.6.0.}
\changes{v1.4.0-1}{2015/05/30}{Fix release: corrected a few typos and errors in the manual.}
\changes{v1.4.0}{2015/05/28}{First public release (version number set to match the included \texttt{academicons.ttf} font version).}

\begin{document}
\title{Analyzing the evolution and diversity of SBES Program Committee}
\author{%
  Fabio Pacheco\thanks{Email: \href{mailto:fabiomelg@gmail.com}{\tt fabiomelg@gmail.com}}, 
  Igor Wiese\thanks{Email: \href{mailto:igors@utfpr.edu.br}{\tt igors@utfpr.edu.br}}, 
  Bruno Cartaxo\thanks{Email: \href{mailto:email@brunocartaxo.com}{\tt email@brunocartaxo.com}},\\
  Igor Steinmacher\thanks{Email: \href{mailto:igor.steinmacher@nau.edu}{\tt igor.steinmacher@nau.edu}}, 
  Gustavo Pinto\thanks{Email: \href{mailto:gpinto@ufpa.br}{\tt gpinto@ufpa.br}} \\
}
\maketitle

\begin{abstract}
The Brazilian Symposium on Software Engineering (SBES) is one of the most important Latin American Software Engineering conferences. It was first held in 1987, and in 2019 marks its 33rd edition. Over these years, many researchers have participated in SBES, attending the conference, submitting, and reviewing papers. The researchers who participate in the Program Committee (PC) and perform the reviewers' role are fundamentally important to SBES, since their evaluations (e.g., deciding whether a paper is accepted or not) have the potential of drawing what SBES is now. 
Knowing that diversity is an important aspect of any group work, we wanted to understand diversity in the SBES PC community.  We investigated a number of characteristics of SBES PC members, including their gender and geographic location. We also analyzed the turnover and renovation of the committee. Among the findings, we observed that although the number of participants in the SBES PC has increased over the years, most of them are men ($\sim$80\%) and from the Southeast and Northeast of Brazil, with very few members from the North region. 
We also observed that there is a small turnover: during the 2010 decade, only 11\% of new members were added to the PC. Finally, we investigated the participation of the PC members publishing papers at SBES. We observed that only 24\% of the papers accepted to SBES were authored by members who were not committee members of the respective year. Moreover, committee members usually do not collaborate among themselves: a significant number of the papers are authored by the PC members and students.
This paper may contribute to the SBES community, in particular, its special interest group, in understanding the needs and challenges of the PC's participants.

\end{abstract}

\bigskip

\section{Introduction}

Unlike other areas of knowledge, in the field of computer science some conferences with rigorous peer review processes are notoriously as well regarded as high impact international journals. Particularly in the field of Software Engineering, the acceptance rate of papers at renowned international conferences is around 20\% to 30\% of the total papers submitted~\citep{VASILESCU2014251}. Conferences also have one advantage: the presentation exposes the work to dozens or even hundreds of stakeholders, potentially increasing its visibility. Thus, the publication and dissemination of scientific papers in Computer Science in general, and Software Engineering in particular, are done not only through journals, but also in conferences.

While journals, which are driven by an editorial committee that includes associate editors, as well as editorial staff, the decision-making in Software Engineering conferences follows a different structure, led by a Program Committee (PC). Rather than having a chief editor and his/her associate editors (who usually support the magazine for a long period), the Program Committee (PC) is set up in advance, and a coordinator (also known as \textit{chair}) of the program committee, who is responsible for, among other activities, dividing the submitted papers amongst committee members for review and fostering discussion to reach consensus on the eventual acceptance/rejection of the papers. Although the names of members of program committee members are usually publicly available, which member reviewed each paper is confidential. With this anonymity, members can more freely exercise their decision-making power.
Although the PC chair formally makes the final decision on whether papers are accepted or rejected, the chair commonly accepts the decision of the PC based on an informal agreement to do so. Thus, in practice, the PC members collectively decide what papers will be presented. In a larger sense, the program committee (and its community) steer the future of that conference. If a significant number of members are strongly active in particular research area, papers addressing this topic will likely be reviewed with greater interest. Similarly, papers that address issues unconnected to the committee's interest may be less favorably reviewed.  Thus, the choice of the program committee is essential not only to balance representation of diverse research topics, but also to include members that representative of diverse genders, localities, tenure, etc.

This paper aims to investigate how diverse is the community of the Brazilian Symposium on Software Engineering (SBES) in terms of its program committee. SBES is the main Brazilian event in Software Engineering. In 2019, SBES had its 33rd edition. Over this period, a few hundred researchers have served as members of the SBES PC. This paper is an extension of a short paper published in SBES in 2019~\citep{Pacheco:SBES:2019}. In the SBES short paper, we initially explored the research questions: (1) how has the number of members participating in the program committee evolved over the years?; (2) what is the geographical distribution of members?; and (3) what is the percentage of female participation on the program committee? However, some relevant questions were not yet addressed, such as: (1) how do SBES program committee members renew?; and (2) how are SBES publications distributed in relation to their committee?

To answer these questions and deepen the discussions previously presented in the SBES short paper, a mostly quantitative study was conducted on the evolution of SBES PC members. Based on SBES event information available from the Special Software Engineering Commission (CEES) website, all researchers who participated as a program committee member between 1998 and 2019 were cataloged.\footnote{We could not find online information before 1998.} Then the data was organized, grouped, and normalized. Combined with previous work~\citep{Pacheco:SBES:2019}, our main findings are described below:

\begin{itemize}
    \item The number of members in the SBES Program Committee has increased over time. However, this increase is mostly driven by national members rather than international members. Forty-Nine researchers (19\%) participated as PC members more than 10 times.
    \item Women's participation is significantly lower than men's (about four male researchers per female researcher). The picture is different when it comes to senior members: of the 34 members who participated in at least half of the editions reviewed (12 editions), 14 are women.
    \item  The Northeast and Southeast regions have consistently shown a larger number of members on the SBES committee. The widening of this disparity was marked in the first decade of analysis ($\sim$1998--2008).
    \item There is relatively little inclusion of new members in the program committee. An analysis of the last 10 SBES editions shows that from year to year 11\% of the committee comprises new members. We found no correlation between program committee size and committee retention.
    \item About 24\% of SBES accepted publications are not authored by committee members. Each year, about 20\% of the committee has publishes in SBES annually. Committee members do not appear to collaborate much with other committee members, suggesting an interaction of their committee members with their student(s) or researchers / industry members who are not part of the committee.
\end{itemize}

\section{Related Works}

This section is organized into bibliometrics work in Software Engineering (Section~\ref{sec:relatedgeral}) in general and SBES in particular (Section~\ref{sec:relatedsbes}).

\subsection{Bibliometric Studies in Software Engineering}\label{sec:relatedgeral}

\cite{VASILESCU2014251} and colleagues discuss the health of Software Engineering conferences based on program committees data and accepted papers data sets. The study defined a set of metrics and noted the program committee renewal rate from one year to the next, and introversion defined as the proportion of accepted papers containing at least one program committee member, are high (above 50\%) at the most reputable conferences (\textit{CORE Rank}\footnote{Ranking internacional de conferências: http://portal.core.edu.au/conf-ranks/}). Overall, the authors found that the lower the retention of committee members, the greater the proportion of accepted papers containing program members. When this occurs, the conferences tend to have a high renewal of authors with papers accepted in relation to the previous edition (70\%), and may reach 50\% of new authors considering the last 4 years. According to the authors, one of the possible reasons is the use of  \textit{ACM SIGSOFT policy}\footnote{https://www.sigsoft.org/policies/pgmcommittee.html} which establishes a renewal of at least 1/3 of the program committee members each year.

\cite{Cabot:2018:CCC:3281635.3209580} and colleagues studied 65 conferences in 2015 and showed evidence that higher ranking conferences are closer to new authors than lower ranking conferences. For example, while the median of novice authors is 14\% in CORE Ranking A*, in CORE Ranking A, B and C is 10\% higher. When satellite events are co-located with the main event, it is clear that these events attract new attendees and, eventually, new authors. This phenomenon was also evidenced by~\cite{VASILESCU2014251} and a possible explanation may be a lower acceptance rate used by conferences with higher CORE Rankings.

In Brazil in recent years, questions have been raised about how the scientific community and events are organized has also gained attention in recent years. \cite{csindexbr} created CSIndexBR, an online platform that provides transparent data on Brazilian Computer Science papers by indexing full research papers published since 2014 by Brazilian researchers at selected conferences and journals. Based on CSIndexBR, the authors also published a recent study on Brazilian author publications at software maintenance and evolution conferences~\citep{MTVem2018}.

\cite{Freire:2018} present a retrospective of the last years of the Brazilian Symposium on Software Quality, indicating the most relevant topics, the most active researchers, as the industry participates in the conference. However, no discussion of how the community is opened has been held. 

The literature also includes studies that seek to understand most used scientific methods  in conferences and journals in computer science; and to analyze co-authoring networks analyzing conference papers~\citep{Zannier:2006,GLASS2002491,LEMOS2013951,Alan,MariaDBLP,MariaSouto}. Although interesting, this is not the focus of this paper.

\subsection{Studies involving SBES} \label{sec:relatedsbes}

\cite{Monteiro:2017} and colleagues conducted a quasi-controlled experiment with 201 papers published between 2007--2016 in SBES. The paper discussed a portrait of which empirical methods were most commonly used, how validations were made, and how threats to validity were reported. The paper made a number of recommendations for researchers in the software engineering area.

\cite{SILVEIRANETO2013872} and colleagues conducted an analysis of the evolution of SBES from the first 24 editions of SBES history compared to the international community,  using ICSE as a benchmark. The authors concluded that the interest of the SBES scientific community has increased and that over the years the acceptance rate is around 25\%.  Regarding community openness, the authors present a single analysis of the relationship of national and international program committee members. In the first editions they found that 37\% of the committee comprised international members, with the majority being British (56), North American (54), Italians (25), Portuguese (23), Canadians (20), Argentinians.

In an initial study, the authors of this paper investigated the evolution of the SBES program committee in terms of: number of participants (national and international) over the years; gender; and location~\citep{Pacheco:SBES:2019}. The present work is an extension of this study.

Despite the studies cited, little has been discussed about how the SBES program committee was formed. As seen in other works, e.g.,~\citep{VASILESCU2014251,Cabot:2018:CCC:3281635.3209580,MTVem2018}, studying the organization of the scientific community and how the program committee is renewed is important to attract new researchers. In the next sections we present our research questions.

\section{Research method}
This section presents the research questions, data collection, analysis procedure, and the replication package. 

\subsection{Research Questions}

This paper aims to answer five research questions. They are:

\begin{itemize}
    \item[\textbf{RQ1}.] How has the SBES program committee evolved over the years?
    \item[\textbf{RQ2}.] How is the women participation in the SBES program committee?
    \item[\textbf{RQ3}.] How are SBES program committee members geographically distributed?
    \item[\textbf{RQ4}.] How does SBES program committee renewal take place?
    \item[\textbf{RQ5.}] How are SBES publications distributed in relation to its committee?
\end{itemize}

In general, these questions explore aspects related to the inclusion of women on the program committee, the geographical distribution of committee members, member renewal (including the departure and entry of new members on the committee), and the distribution of publications in relation to the members of the committee. We believe such research questions are relevant to better understand the SBES community, since, if women's participation is low, or few members in a particular locality are represented, efforts may be made to increase participation of these groups.  Finally, answers to these research questions may enable discussion about the overall diversity of the program committee. These issues are often discussed at major software engineering conferences around the world, as seen earlier in related work

Although revisited and deepened in the present work, the first three research questions were initially presented in the work of~\cite{Pacheco:SBES:2019}. The last two questions are unique to this paper.

\subsection{Data collection}

To collect information from members of the SBES program committee, we consulted the web-sites of the previous editions. The CEES\footnote{http://comissoes.sbc.org.br/ce-es/} website has a table indicating the events that took place, from the first in 1987, in Petrópolis, Rio de Janeiro, to the last, in 2018, in São Carlos, in the interior of São Paulo. However, only pointers remain to the websites that occurred between 1998 and 2019. Thus, the first 11 SBES editions no longer have their websites available for consultation through the CEES website.  In addition, some of the event sites from 1998 to 2018 yield 404 errors (page not found). 

For some of these cases, this issue could be mitigated by using the \textit{Internet Archive} (\texttt{http://web.archive.org/}) website. The \textit{internet Archive} is a non-profit organization that aims to build a digital library of websites and other digital artifacts. 
Using this site, it was possible to retrieve 11 instances of old SBES websites that were otherwise unavailable. Unfortunately, however, even with using \textit{internet Archive}, it was not possible to retrieve information from the 2002 edition. Thus, this research covers the period from 1998 to 2019, except for 2002, totaling \numEdicoes SBES editions.

\subsubsection{Data Normalization}

After collecting data from committee members, it was observed that throughout the editions there was no standardization regarding the name of the members or the name of their institutions. For example, the PC member ``Itana Maria de Souza Gimenes" in some editions has been described as ``Itana Gimenes" or ``Itana M. de Souza Gimenes". These occurrences were normalized to "Itana Maria de Souza Gimenes". Similarly, this particular member is affiliated with the Maringá State University. Throughout the editions, this institution has been described as ``DI-UEM", ``UEM-PR", ``DI/UEM/PR", or simply ``UEM". In this particular case, the institution was normalized to UEM. The normalization process has been applied to all members and all their institutions.

Moreover, during the data analysis, researchers were distinguished as national (those affiliated with Brazilian institutions) or international (those af- filiated with institutions abroad). This data normalization process was conducted by the first author with the support of the last author. Ultimately, we found that \totalMembros unique members participated in the SBES committee at least once.



\subsection{Data analysis}


For data analysis, we followed a mostly quantitative approach, supported by descriptive statistics techniques. For example, we used bar graphs with absolute numbers, time series line graphs, and trend measures (e.g. mean, median, interquartile ranges, etc.).

\subsection{Replication Package}

All data reported in this work is available online in a Google Drive spreadsheet for consultation and replication at the following address: \texttt{\url{http://bit.ly/sbesPC}}.

\section{Results}

Results are organized in terms of research questions.

\subsection{RQ1. How has the SBES Program Committee evolved over the years?}

\begin{figure}[!ht]
\begin{center}
 \includegraphics[width=\columnwidth]{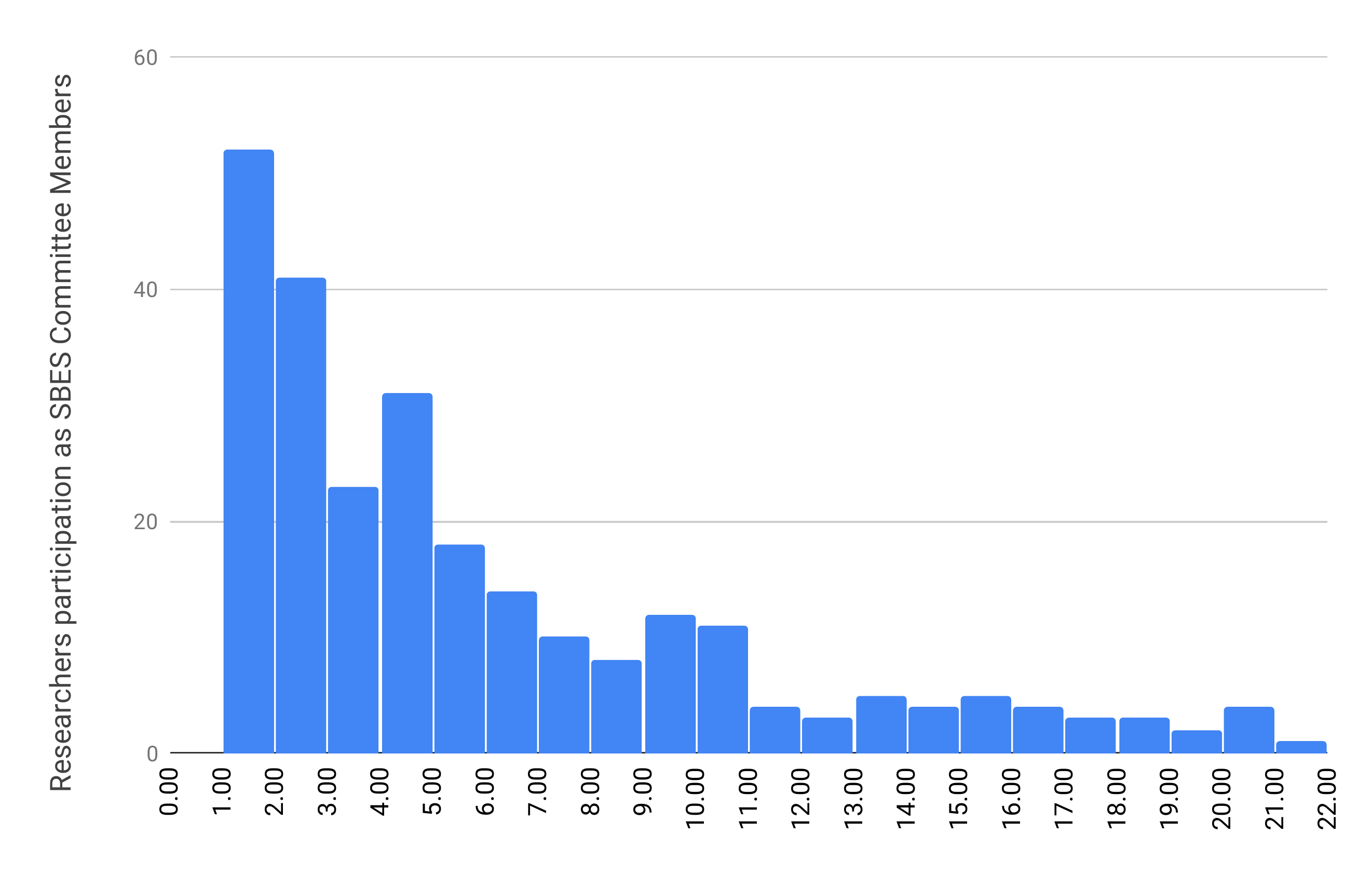}
\end{center}
\caption{Histogram summarizing the number of instances that each committee member participated}
\label{fig:hist_participacao}
\end{figure}

Initially, we found that \totalMembros researchers were part of the SBES committee over the \numEdicoes symposium editions mapped by this study. These researchers have a median of 4 participation instances in the SBES PC (Q1 = 2, Q3 = 8). Figure~\ref{fig:hist_participacao} shows the distribution of committee members' participation. Most researchers participated as members of the SBES PC between one and four times (116/46\%), while 49 researchers (19\%) participated in the SBES PC during 10 or more editions. 

\begin{figure}[!ht]
\begin{center}
 \includegraphics[width=\columnwidth]{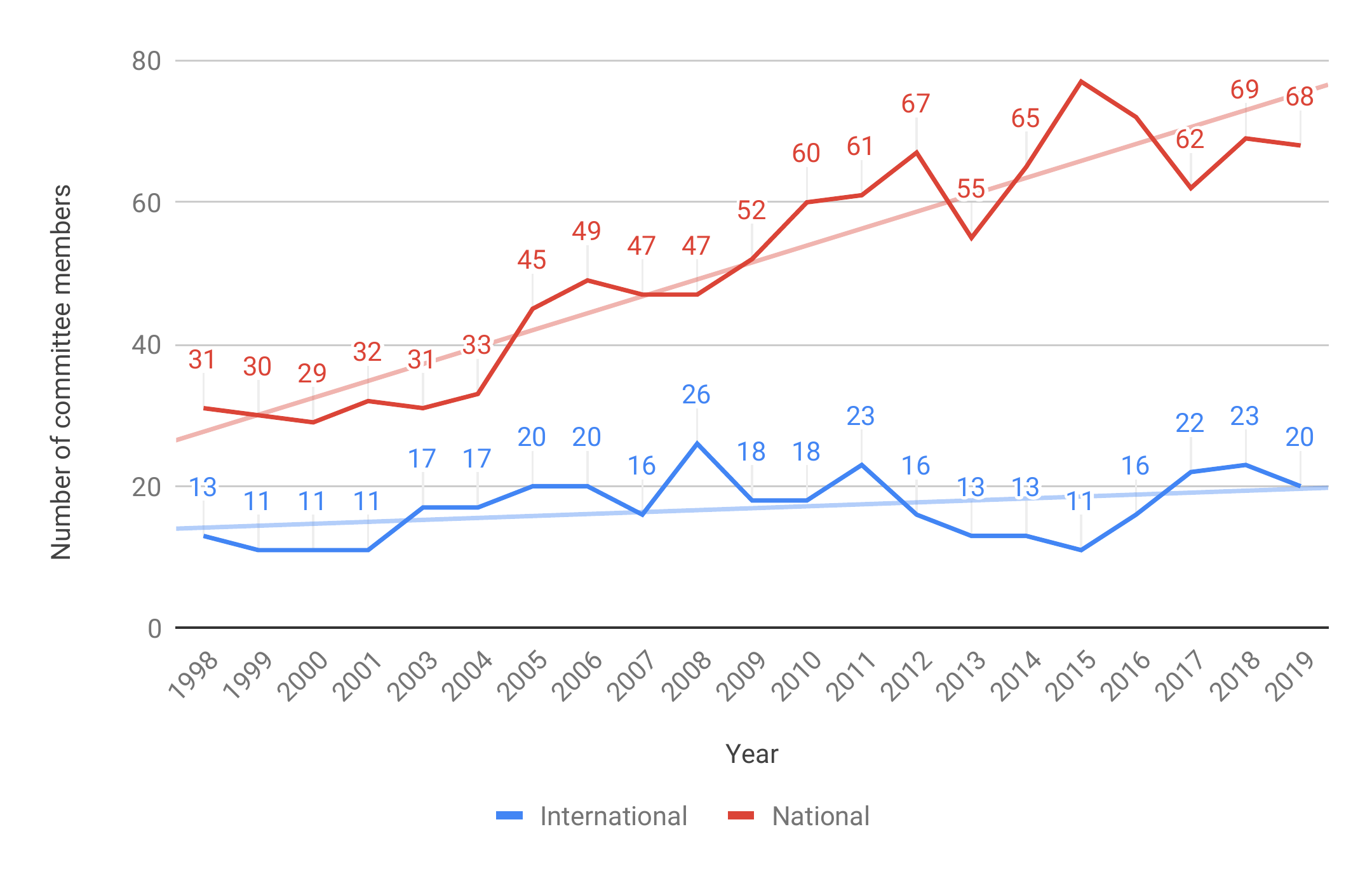}
\end{center}
\caption{Number of SBES Program Committee members, split in national and international members.}
\label{fig:evolucao_nacional_internacional}
\end{figure}

A more detailed view is presented in Figure~\ref{fig:evolucao_nacional_internacional}. The top line (in red) indicates the number of researchers from Brazilian institutions who participated as members of the SBES program committee, while  the bottom line (in blue) indicates the number of researchers from international institutions. For each, another softer-toned line indicates the trend of the main line

As can be seen from this figure, the number of Brazilian researchers who joined the SBES PC has increased over time, suggesting linear growth along the trend line. In the first data record, there were 32 national researchers on the program committee. By 2019, there were 72 national members (a growth of $ 2.25\times$, when compared to 1998). The participation of international members, although growing overall, grew at a much more modest pace compared to national members (from 13 members in 1998 to 20 members in 2019). 

Overall, there were never fewer than 11 researchers from international institutions and 30 researchers from national institutions. The year with the largest number of researchers was 2018 with 92 committee members (23 international and 70 national), while 2000 had 41 members (11 international and 30 national) representing the smallest SBES PC in the history of the symposium. Interestingly, it took the committee nearly 20 years to double its size. 
Interestingly, it took the committee nearly 20 years to double its size.  Although the median size of the SBES PC is 71 (Q1 = 50, Q3 = 84), it was not until 2006 that the symposium had more than 70 members on its program committee. While there was a sudden drop in committee members from 83 members in 2012 to 68 in 2013, in the following year 10 new members reentered on the research group.

\MyBox{\textbf{RQ1 Summary}: The number of members in the SBES PC is increasing over the years. This increase is mostly driven by national members. A significant portion of the researchers (46\%) participated as committee members no more than four times. Only 49 researchers (19\%) participated as committee members more than 10 times.}

\subsection{RQ2. How is the women participation in the SBES program committee?}

\begin{figure}[!ht]
\begin{center}
\includegraphics[width=\columnwidth]{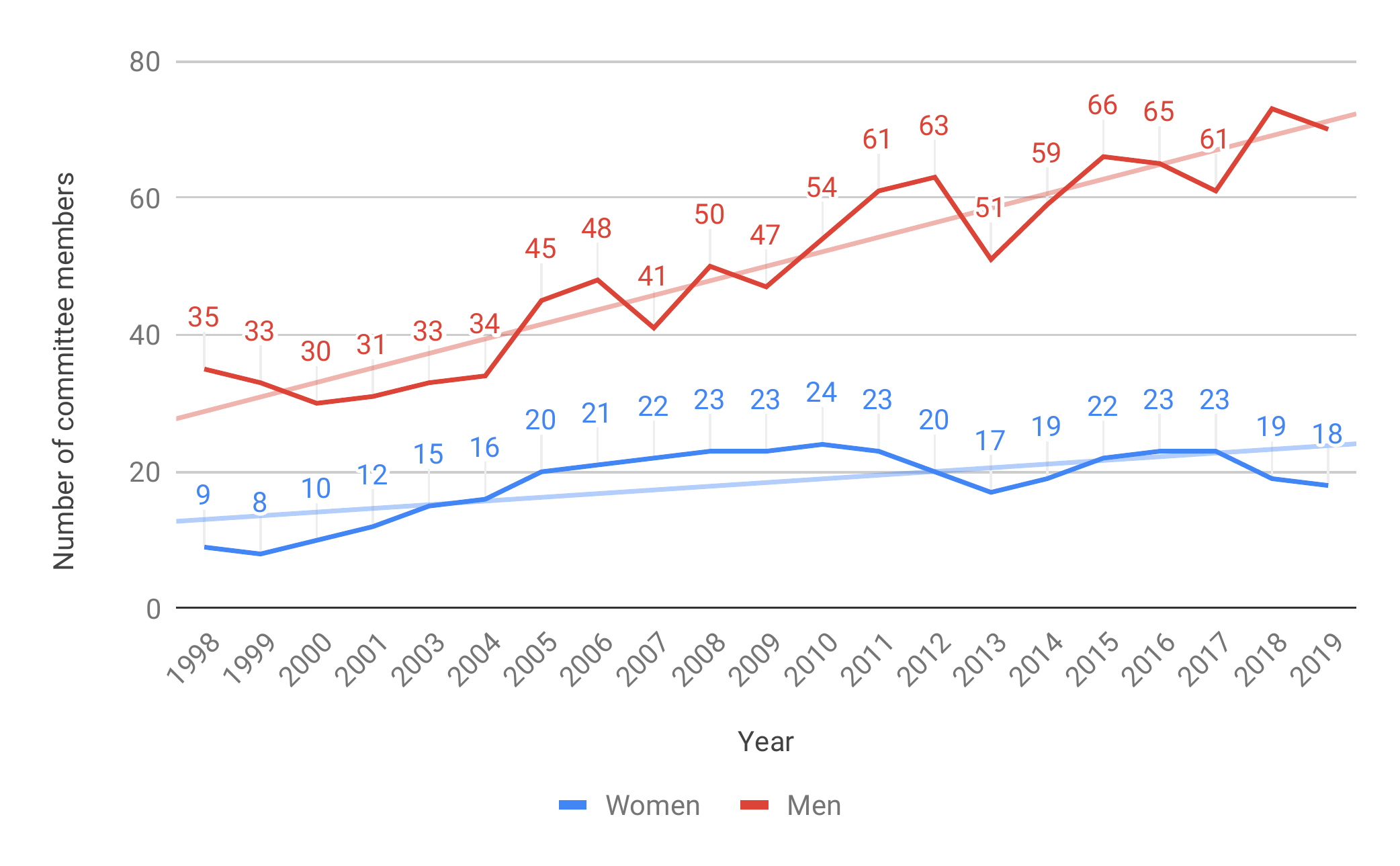}
\end{center}
\caption{SBES PC evolution over the years by gender.}
\label{fig:evolucao_mulheres}
\end{figure}

In recent years, there has been discussion about the difficulties women face in the software community~\citep{TerrellKMRMPS17,IzquierdoHSR19,VasilescuPRBSDF15,Vivian:Cyberfeminist}, and the barriers to enrolling in technology courses~\citep{pintotraining}. Based on this premise, we investigated how open is the SBES program committee is regarding the inclusion of women. Figure~\ref{fig:evolucao_mulheres} shows the evolution between the participation of men (red line) and women (blue line).

Firstly, the figure highlights how male participation was always greater than female participation.  In the first year of this study (1998), there were 9 women and 35 men (3.8 male researchers for each female researcher).  Throughout the years, despite other changes, the ratio of men/women in the SBES PC remained basically the same.  In the latest SBES edition (2019), there were 18 women and 70 men (again, 3.8 male researchers for each female researcher).
Moreover, since the first SBES PC data available (1998), women have always been part of the program committee. At the time of 1998, nine women were part of the committee, eight Brazilian and one international. In 2019, the participation of women doubled, totaling 18 women, of which only three were affiliated with international institutions.

Finally, we conducted an analysis regarding members seniority. When we compared women's participation, taking into account program committee members who participated in at least 12 editions on the SBES program committee --- which is half the editions we analyzed --- we observed that there are 14 women and 20 men. When the comparison is made taking into account the 10 members who most participated in the program committee over the years analyzed, we observed that three members were women. 

\MyBox{\textbf{RQ2 Summary:} In all SBES editions women have participated in the program committee. However, female participation is significantly lower than male participation (about four male researchers for each female researcher). However, the figure differs when analyzing senior members:  of the 34 members who participated in at least half of the SBES edition analyzed (12 editions), 41\% are women.}

\subsection{RQ3. How are SBES program committee members geographically distributed?}

\begin{figure}[!ht]
\begin{center}
 \includegraphics[width=.7\columnwidth]{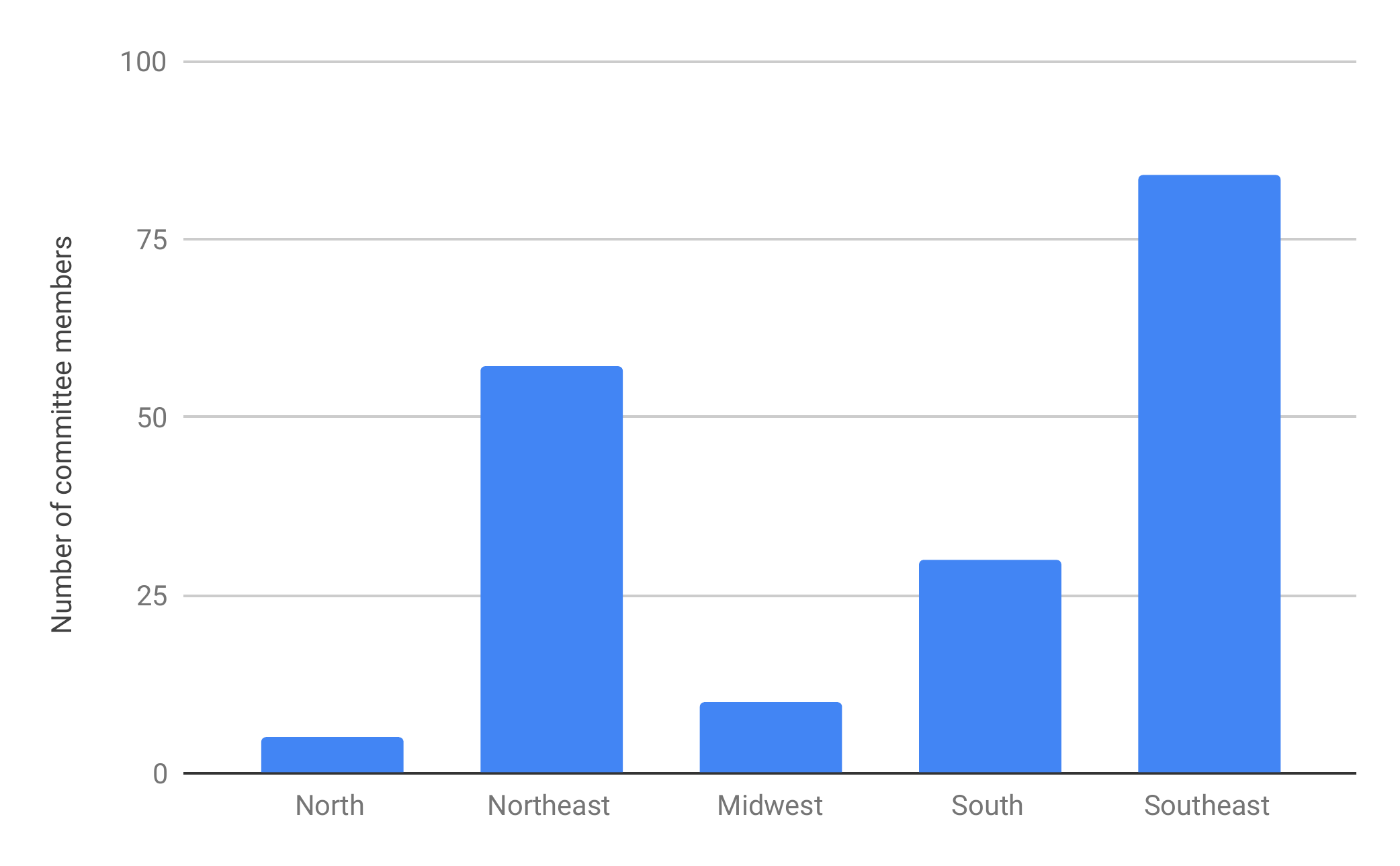}
\end{center}
\caption{Overall number of all Brazilian members who have been part of the SBES committee, grouped by Brazilian geographic regions.}
\label{fig:quantitativo_regioes}
\end{figure}

Figure~\ref{fig:quantitativo_regioes} shows the number of all Brazilian members who have already participated the SBES committee, organized by their geographic location. As you can see, the southeastern region has the largest number of members (84), while the northeastern region has the second largest number of members (57).
The northern and midwest regions have the lowest absolute number of committee members at five and ten, respectively. 

The disparity between geographic location may be related to the mere demographic characteristic of researchers in these locations (e.g., if there are a large number of researchers in one region, proportionally to the others, it is natural that there are more members of that region), or it may be a sign of a lack of representativeness (e.g., if there are many researchers in a location but few of them participate in the SBES PC). To provide more evidence along these lines, a demographic survey of researchers and software engineering research institutions in each region is required. In Section~\ref{sec:discussao}, we present such an initial comparison based on data collected from CSIndexBR.

Figure~\ref{fig:quantitativo_regioes} shows the year-on-year evolution of the number of Brazilian SBES committee members stratified across the five macro regions of Brazil. The first decade analyzed shows a clear growth in the number of members of the Northeast and Southeast regions, followed by a trend of stabilization of this number in the following decade. As for the North, Midwest, and South regions, the number remained stable (with values significantly lower than the Northeast and Southeast regions) throughout the analyzed period. One small exception is the southern region, which has slightly increased over the last five years of analysis. This increase is mainly observed in the number of researchers from Paraná, which jumped from 2 to 9 between 2015 and 2019 (5 of them from UTFPR, which was not represented before 2017).

\begin{figure}[!ht]
\begin{center}
 \includegraphics[width=\columnwidth]{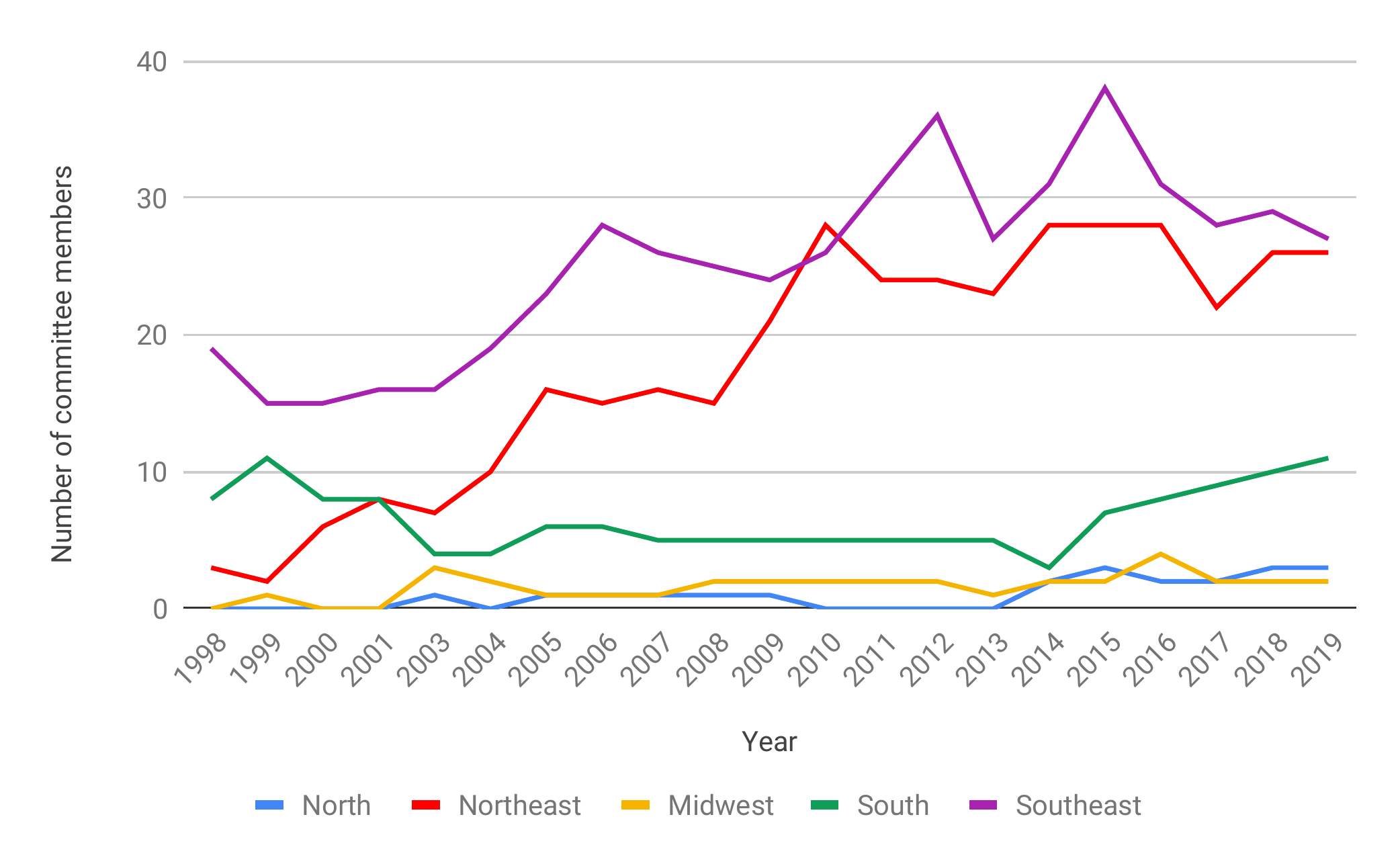}
\end{center}
\caption{Yearly evolution of the number of SBES Program Committee members by Brazilian geographic regions.}
\label{fig:evolucao_regioes}
\end{figure}

The disparity in the trend of evolution of the Northeast and Southeast regions in relation to the other regions further accentuated the difference in representativeness of SBES committee members in relation to the Brazilian regions. However, as mentioned earlier, this disparity may have occurred due to the demographic characteristics of the regions. In section \ref{sec:discussao}, we present such a discussion based on data collected from CSIndexBR.




\MyBox{\textbf{RQ3 Summary}: The Northeast and Southeast regions have consistently presented a larger number of members on the SBES committee than the three other regions of Brazil. The widening of this disparity was marked in the first decade of analysis (\textasciitilde1998–2008).} 

\subsection{RQ4. How does SBES program committee renewal take place?}

\begin{figure}[!ht]
\begin{center}
 \includegraphics[width=\columnwidth, clip=true, trim= 0px 0px 0px 0px]{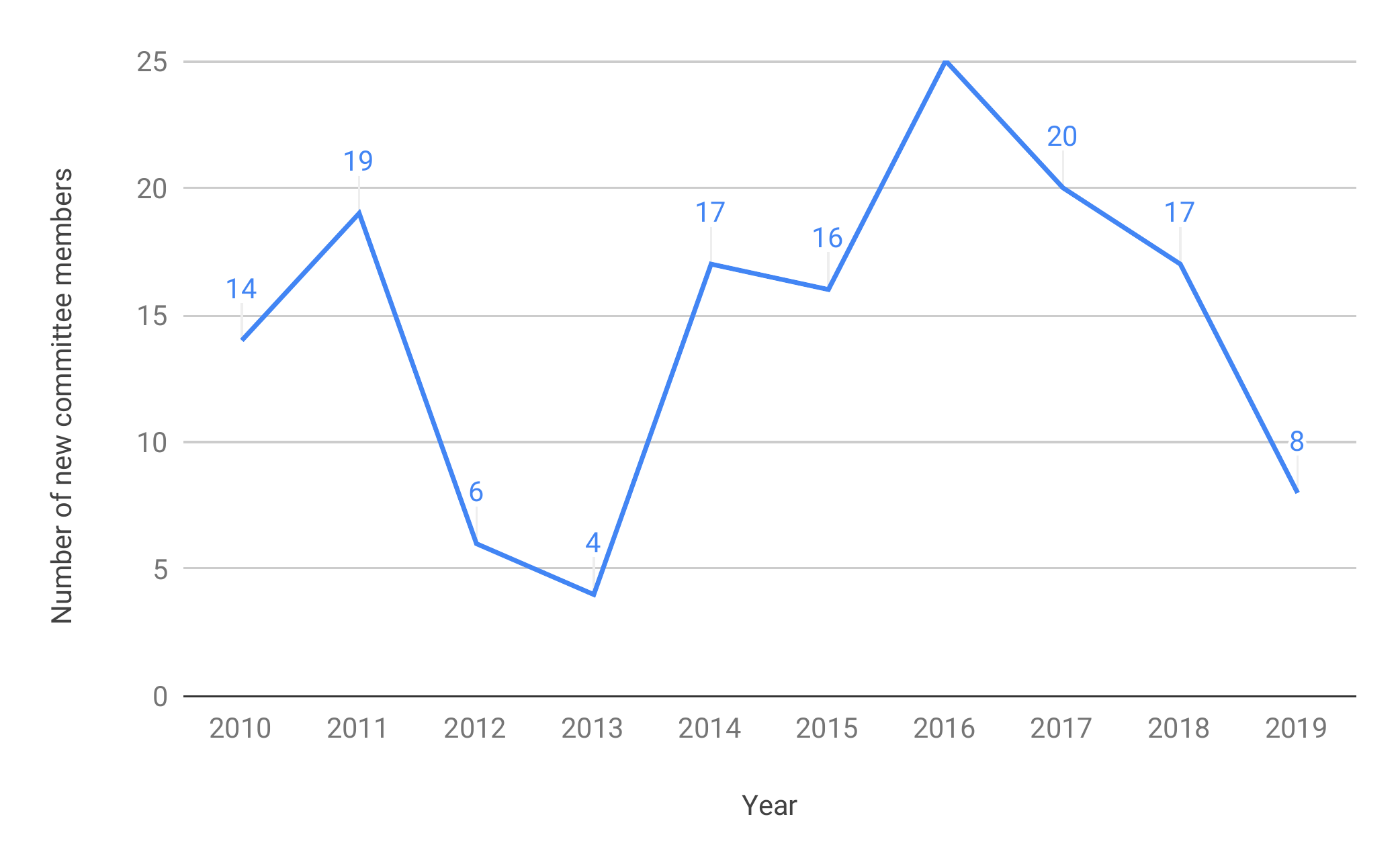}
\end{center}
\caption{ Evolution of new committee members in the last 10 years.}
\label{fig:rotatividade}
\end{figure}

We conducted a more detailed analysis of program committee member turnover with data from the last 10 years of the symposium. Pair by pair, we analyzed how many new members joined the committee from one year to the next. Figure~\ref{fig:rotatividade} presents the results. As can be observed, there was a peak of renewal in the committee in 2016, followed by 2017, 2018 and 2011. It is clear that in 2019 there was little addition of new members; only 9 new members were added this year. 

When calculating the new member rate, it was found that 2017 (25\%) and 2016 (29\%) saw the greatest renewal proportional to the number of members in the program committee. In the years of 2010 (18\%), 2011 (22\%), 2014 (23\%) and 2015 (18\%) this value approached 25\%, indicated as healthy according to other works (e.g.,~\citep{VASILESCU2014251}). However, it was also possible to notice that in 2012, 2013, and 2019, these values were less than 11\% of renovation. Still regarding renewal, it was found that on average 1/4 involve members of international institutions. 

\begin{figure}[!ht]
\begin{center}
\includegraphics[width=\columnwidth]{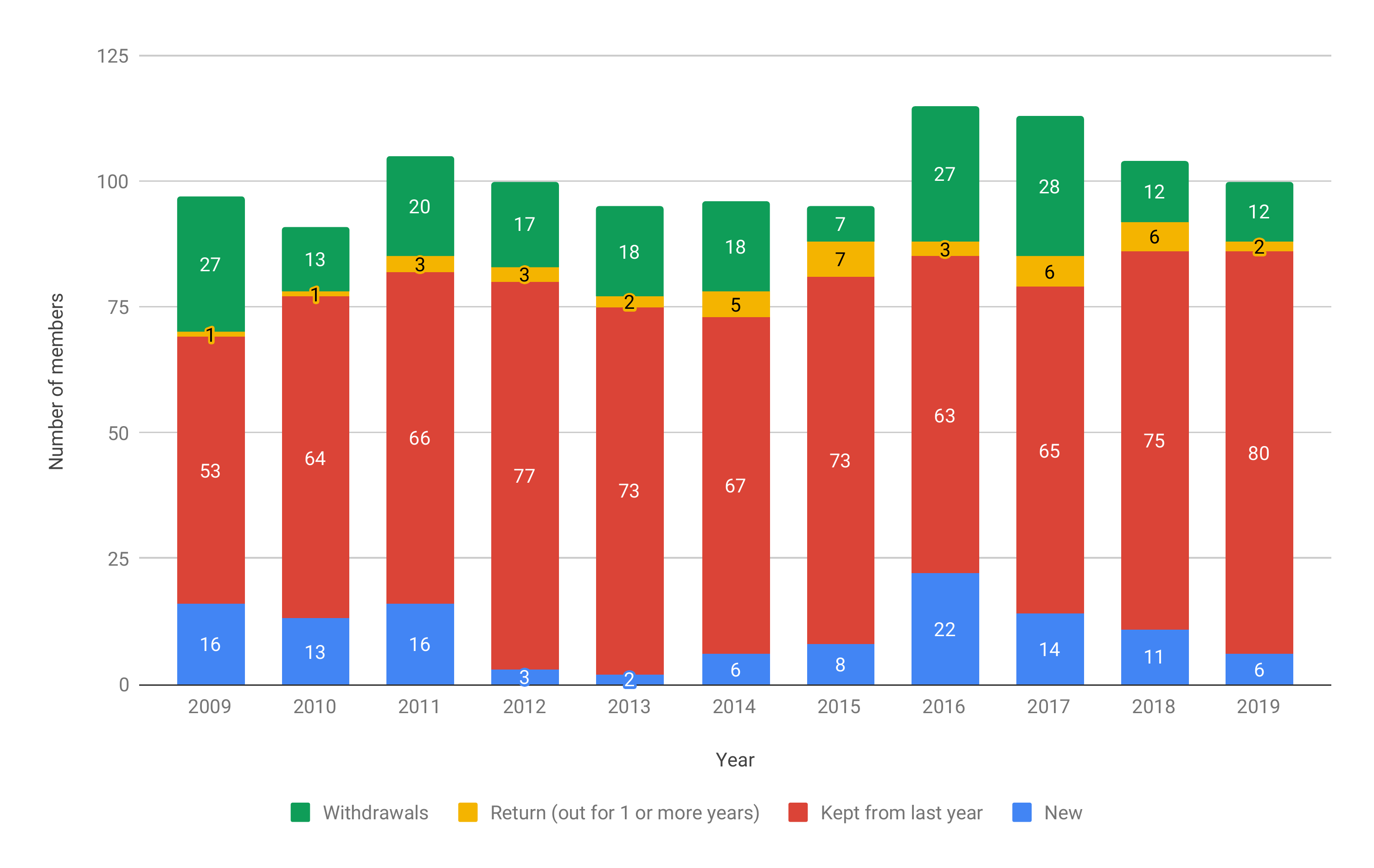}
\end{center}
\caption{Evolution of Program Committee Members: Withdrawals, returning members, members kept, and new members.}
\label{fig:turnover}
\end{figure}

Figure~\ref{fig:turnover} gives a more detailed view of program committee renewal, taking into account each year's committee compared to the previous year's. We looked at the members who remained with the previous year, those who are returning after one or more years out of the committee, those who are new (appearing for the first time), and those who left (were members in the previous year and not the year under review). As you can see, in 8 of the 11 years represented, the membership (or return) of members on the committee is less than the number of members who left. Exceptions are the years 2010, 2015 and 2018; however, these are the years with the lowest number of departures (13, 7, and 12, respectively).

Still on Figure~\ref{fig:turnover}, we can see that the highest committee renewal rate, with the entry of new members, occurred in the years when we observed the highest number of committee exits (2009, 2011, 2016 and 2017). These years coincidentally saw the four highest new member entry rates, including those returning.

An interesting phenomenon occurred in 2016 and 2017, when a major renovation apparently took place in two consecutive years. In these two years, a there were a total of 36 new members. However, by analyzing the data more carefully, we realized that in 2017, 12 of the 22 members included in 2016 were withdrawn in 2017 (43\% of members leaving in 2017 had their first participation in 2016). So even though it looks like a major renewal, if we compare the 2015 committee with the 2017 committee, we realize that 58 members remained, with the actual renewal in these two years being 24 new members and three returning to the committee.

\begin{figure}[!ht]
\begin{center}
\includegraphics[width=\columnwidth]{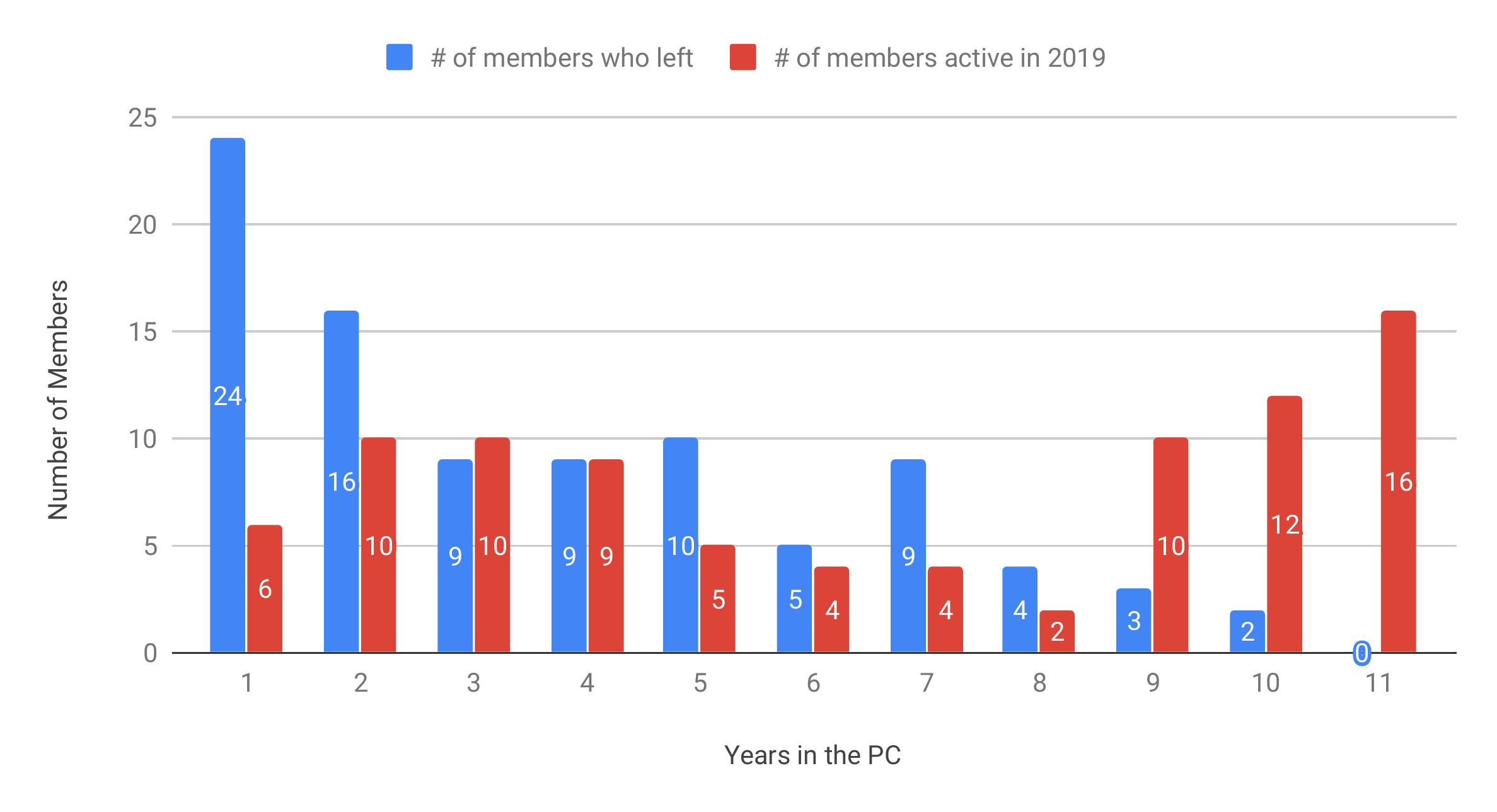}
\end{center}
\caption{Life span of SBES committee members.}
\label{fig:tempoDeVida}
\end{figure}

This observation led us to seek to understand the life span of committee members. The group containing all members who passed through the committee between 2009 and 2019 was analyzed: 179 distinguished researchers. We divided these researchers into those who remain on the committee until 2019, and those who left sometime between 2010 and 2019 (inclusive). Figure~\ref{fig:tempoDeVida} shows the number of appearances of members of each group over the last 11 years. The first observation to make is that 91 of these researchers (51\%) are not part of the 2019 committee. In addition, 44\% of researchers do not serve more than 2 years on the program committee before leaving (24 left with 1 or 2 years on the committee). This data that indicates that the retention rate of new members is low, and needs to be observed. 

Still in Figure~\ref{fig:tempoDeVida}, observing the members that remain until 2019 (red bars), it can be seen that most of the current committee members have had a long tenure, with 38 of the 88 members (43\%) having served at least 9 years on the program committee.  This shows that a group of experienced researchers have maintained the senior character of the committee.


Finally, we calculated the Pearson correlation between program committee size and renewal and found a moderate positive correlation (0.56), which leads us to believe that a larger committee is not always synonymous with renewal. This was also clear in the evolution presented in Figure~\ref{fig:turnover}.

\MyBox{\textbf{RQ4 Summary}: There is little turnover of program committee members if we consider the average number of members that remain on the committee. During the 11 years analyzed, only 11\% of new members were included.}

\subsection{RQ5. How are SBES publications distributed in relation to its committee?}

Finally, this research question aims to understand how SBES publications are distributed in relation to the program committee. 

Table~\ref{tab:publicacoes} presents data from the last 10 editions of SBES comparing the number of publications containing at least one program committee member to publications made without any committee members. To perform the analysis, the committee members of each year were compared with the papers accepted in the same year. As can be seen, the publications during 2013 and 2016 mostly contained committee members (only 1 without members). By contrast, in 2009 and 2011 the percentage of publications accepted without committee members exceeded 1/3. It is still noticeable that, in general, 26\% of publications involve authors who are not part of the committee. 

\begin{table}[!ht]
\caption{Number of SBES papers with and without program committee members.}
\centering
\begin{threeparttable}
\begin{tabular}{|l|c|c|c|}
\hline
\multicolumn{1}{|c|}{\textbf{Year}} & \textbf{\begin{tabular}[c]{@{}c@{}}\# Pub. \\ without committee\end{tabular}} & \textbf{\begin{tabular}[c]{@{}c@{}}\# Pub. \\with Committee\end{tabular}} & \textbf{\begin{tabular}[c]{@{}c@{}}(\%) Pub. \\without committee\end{tabular}} \\ \hline\hline
2009  & 11   & 13 & 45,8\%  \\ \hline
2010  & 3  & 16 & 15,7\% \\ \hline
2011   & 12 & 23  & 34,2\%   \\ \hline
2012   & 7 & 17 & 29,1\%    \\
\hline
2013   & 1 & 16 & 5,8\%     \\
\hline
2014   & 4 & 13  & 23,5\%    \\
\hline
2015   & 6 & 15  & 28,5\%    \\
\hline
2016   & 1 & 14  & 6,6\%    \\
\hline
2017   & 8 & 22 & 26,6\%    \\
\hline
2018   & 4 & 12  & 25,0\%     \\
\hline
\end{tabular}
\end{threeparttable}
\label{tab:publicacoes}
\end{table}

Table~\ref{tab:comite} shows how many committee members had papers accepted in each SBES year from 2009. Only 2011 shows a large participation of the committee in relation to the published papers; interestingly, also in 2011 we noticed a high number of papers without any committee member (see Table~\ref{tab:publicacoes}). This suggests that, in 2011, many committee members collaborated to submit papers to the conference.

Interestingly, over these 10 years, there were 161 publications with commit- tee members that were published by 170 members.   This relationship suggests that committee members do not collaborate with each other since few committee members concentrate on accepted submissions with program committee members. On average, over the last 10 years 20\% of the committee has recurrently accepted papers in SBES. 

\begin{table}[!ht]
\caption{Number of committee members with papers at SBES per year}
\centering
\begin{threeparttable}
\begin{tabular}{|l|c|c|c|}
\hline
\multicolumn{1}{|c|}{\textbf{Year}} & \textbf{\begin{tabular}[c]{@{}c@{}}\# Members \\with Pub\end{tabular}} & \textbf{\begin{tabular}[c]{@{}c@{}}\# Members \\without Pub\end{tabular}} & \textbf{\begin{tabular}[c]{@{}c@{}}(\%) Members \\with Pub.\end{tabular}} \\ \hline\hline
2009  & 16   & 54 & 22,8\%  \\ \hline
2010  & 14  & 64 & 17,9\% \\ \hline
2011   & 38 & 46  & 45,2\%   \\ \hline
2012   & 20 & 65 & 23,5\%    \\
\hline
2013   & 14 & 64 & 17,9\%     \\
\hline
2014   & 12 & 67  & 15,1\%    \\
\hline
2015   & 6 & 84  & 6,6\%    \\
\hline
2016   & 14 & 76  & 15,5\%    \\
\hline
2017   & 17 & 68 & 20,0\%    \\
\hline
2018   & 19 & 75  & 20,2\%     \\
\hline
\end{tabular}
\end{threeparttable}
\label{tab:comite}
\end{table}

\MyBox{\textbf{RQ5 Summary}: Around 24\% of SBES accepted publications are not authored by committee members. About 20\% of the committee has published in SBES annually. Committee members do not appear to collaborate much with other committee members when it comes to accepted papers on SBES. This may mean that committee members interact  with their student(s) or researchers/industry members who are not part of the committee.}

\section{Discussion}
\label{sec:discussao}

This section presents the main findings of this paper and discusses them in more detail.

\vspace{0.2cm}
\noindent
\textbf{Does the SBES program committee look at the Brazilian software engineering community well?} In this work, a total of \totalMembros researchers who were member of SBES committee were observed.  A relatively low turnover rate was observed among committee members. Thus, it is possible that the SBES committee chairs have difficulty recruiting new members. In order to compare the ''coverage' of the SBES committee's list of participants, CSIndexBR~\citep{csindexbr} is used as a baseline. CSIndexBR is a platform that indexes contributions made by researchers linked to Brazilian institutions in high impact journals and conferences in the most diverse areas of Computer Science. All Brazilian software engineering researchers indexed in CSIndexBR were considered in this research.  Compared to the data from this work, 78 out of 169 (46\%) Brazilian researchers indexed by CSIndexBR have never participated in the SBES committee (since 1998).  Thus, there is ample opportunity for new members to join the SBES committee

\begin{figure}[!ht]
\begin{center}
 \includegraphics[width=0.7\columnwidth]{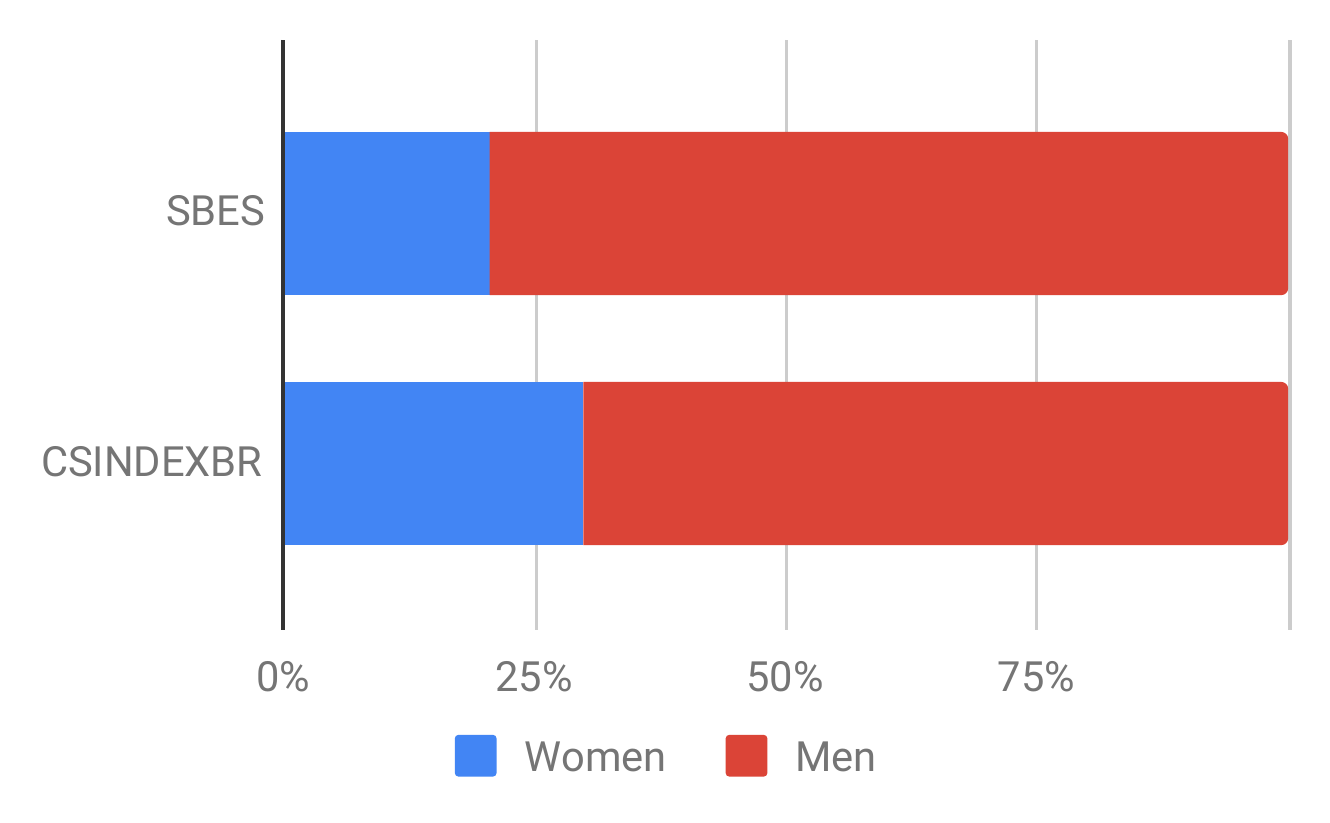}
\end{center}
\caption{Comparing the ratio of men and women on SBES committee and CSIndexBR.}
\label{fig:mulheres-csindex}
\end{figure}

\vspace{0.2cm}
\noindent
\textbf{Could the participation of women in the SBES program committee improve?} 
Although the number of researchers participating in the SBES committee has grown, the difference between the number of men and women remains the same as it was 20 years ago, with about 4 men for each woman. Another point to consider is that recently (2014-2019), only 11 new members joined the committee in relation to 89 new members. Last year, no new members were invited. For comparison purposes, CSIndexBR was also used to gain a sense of the participation of indexed women in the platform. Figure~\ref{fig:mulheres-csindex} compares the comparison between the proportion of women who have already participated in the SBES committee to the women indexed by CSIndexBR, showing a slight similarity between the datasets (20.5\%
of women on the SBES committee and 29.9\% of women on the CSIndexBR). This result suggests that the SBES committee is not necessarily biased towards male participation.

However, of the 78 researchers who have never participated in the SBES committee (discussion above), 28 are women. Because CSIndexBR indexes high-impact\footnote{https://csindexbr.org/faq.html} conferences and journals, this list of researchers offers ample opportunity to improve the committee's renewal rate, including closing the gap between the number of men and women on the committee.

\begin{figure}[!ht]
\begin{center}
\includegraphics[width=0.7\columnwidth]{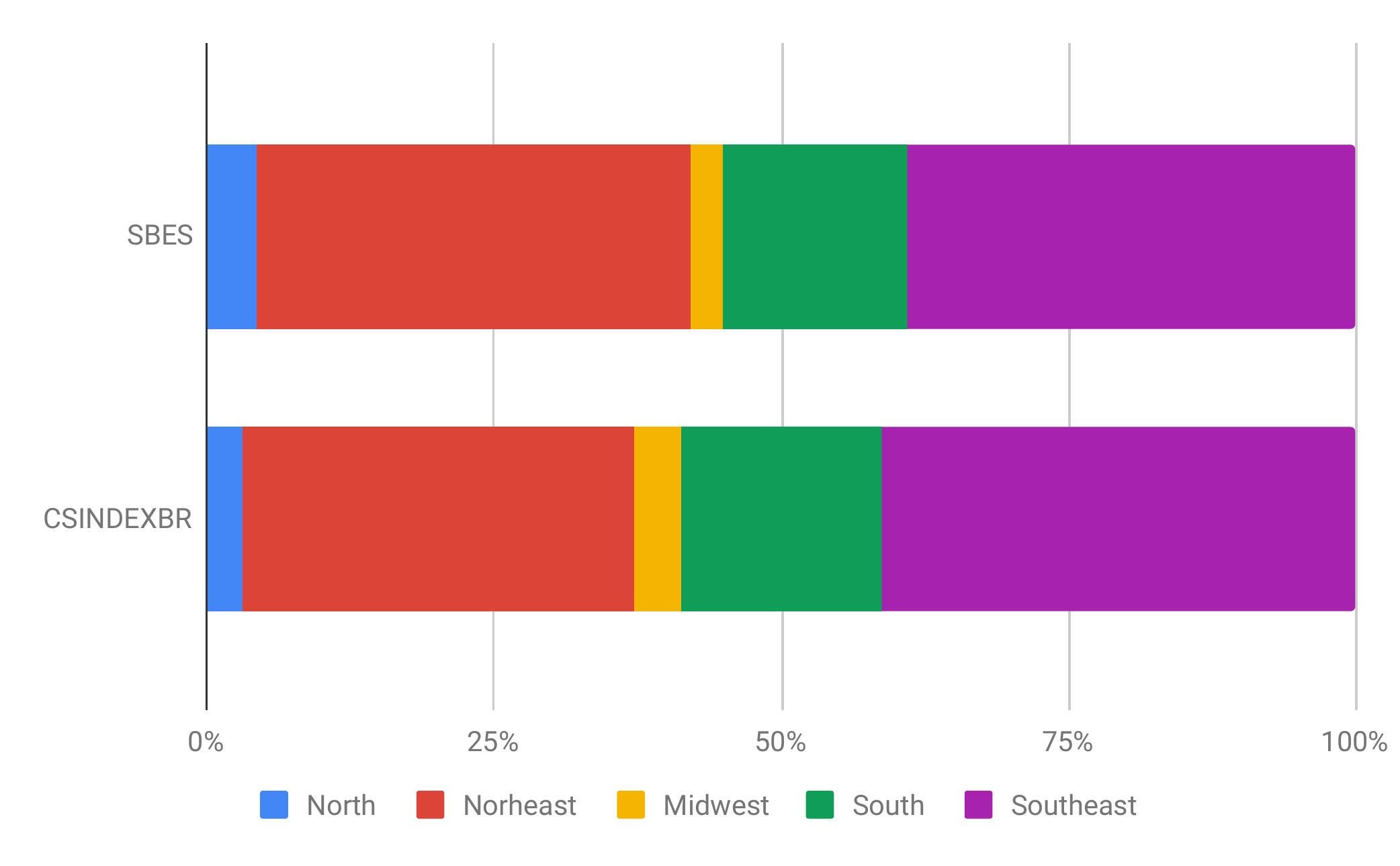}
\end{center}
\caption{Distribution of member participation in SBES committee per region: SBES vs. CSIndexBR.}
\label{fig:regioes-csindex}
\end{figure}

\vspace{0.2cm}
\noindent
\textbf{Is the geographical distribution of SBES members biased?} As noted in RQ3, a significant portion of the members (75\%) who had already participated in the SBES program committee are concentrated in the Northeast (30\%) and Southeast (45\%).  An avid reader might imagine that there might be a certain inclination to include members regarding their demographic region (eg. a Northeast \textit{chair} could more easily invite his colleagues from the Northeast as well). In order to better understand this phenomenon, the CSIndexBR platform was again used to categorize indexed researchers with respect to geographic region. The result is shown in Figure \ref{fig:regioes-csindex}.

As it is possible to notice in the figure, the distribution of members according to the region they are from are similar when comparing the SBES committee and the researchers indexed by the CSIndexBR. This finding indicates  that there is no geographical bias in the SBES committee's participation.

Regarding the internationalization of SBES, the event has been working on initiatives to gain greater international projection and, consequently, greater scientific impact. In particular, the SBES committee has been encouraging, at least in its last 10 editions, that papers should be written in English, although they still accept those written in Portuguese. This can be seen in the following excerpt from the SBES 2019\footnote{\url{https://cbsoft2019.ufba.br/\#/sbesresearchtrack}} \textit{Research Track} paper call page: \textit{``Submission in English is strongly encouraged [...]''}. However, the proportion of international members who joined the program committee is relatively small. Apparently this has diminished, because there is a substantial increase in publications with authors affiliated to international~\citep{SILVEIRANETO2013872} institutions, even if the authors are actually Brazilian affiliated to international institutions.

\vspace{0.2cm}
\noindent
\textbf{Renewal Policy} Internationally, there is much debate about the adoption of policies such as those in \textit{ACM SIGSOFT}~\citep{VASILESCU2014251},  which suggests a minimum renewal of 30\% of the committee from one year to the next. Although we are close to this recommendation (see results in RQ4), this does not occur every year, or even when it occurs the PC actually included of members who had already participated before. The argument in favor of adopting the renewal policy is that encouraging the submission of more papers brings in a new body of authors and research topics. 

\section{Limitations}

Data collection was manually performed. Some implications may occur due to that fact, for example, the number of publications with and without committee members, or the institutions and full names of the co-authors of the SBES papers. To mitigate these limitations, these tasks were verified by more than one co-author of this paper. In addition, the analysis was conducted solely based on results obtained with descriptive statistics, based on graphs, averages, and distributions. While other statistical methods for group comparison could have been used, we determined that descriptive statistics were sufficient to answer the proposed research questions.

The analysis performed on research questions 4 and 5 took into consideration only the editions from 2009 to 2019. We made this cut because we believe the community would be most interested in evaluating recent history, and that this scope would be sufficient to appropriately respond to the RQs.

Some international members are researchers who have made careers in Brazil, but at some point have stepped aside to take another position abroad. This type of behavior generated duplicity in some records, since these researchers were at one point considered international (when in foreign institutions) and at another point national (when in Brazilian institutions). To keep the data in the most reliable format possible, we decided not to merge these researchers.

Finally, when the published papers were analyzed, the names of the authors were raised in the proceedings published in the digital libraries. In some years, however, the proceedings included the track of emerging results, for example. As the purpose of the question was to analyze the committee members' participation in the publications, we decided to consider all the papers presented in the proceedings, even if they came from different tracks.

\section{Conclusion}

The purpose of this paper was to investigate how open is the community of the Brazilian Software Engineering Symposium (SBES), here represented by the members of its program committee. We shed light on how the program committee has historically evolved, specially regarding female participation, geographical distribution of researchers, and how the committee members are renewed.

Among the main findings of this work, the following stand out: 
\begin{itemize}
    \item The number of SBES program committee participants is increasing;
    \item In all editions of SBES, male researchers have been well-represented on the program committee. Unfortunately, however, female participation is significantly lower than male participation (about four male researchers for each female researcher);
    \item We still have few committee representatives in the north and midwest regions;
    \item Member retention is still high from year to year. Few new members have been added over the years, meaning that women are consistently under-represented;
    \item Less than 1/3 of the publications are from non-program committee members.
\end{itemize}

We hope the results will help to reflect on how open the software engineering community is and what actions could be taken to promote greater inclusion, as SBES is the leading software engineering conference in Brazil, and one of most important in Latin America.



\bibliographystyle{ACM-Reference-Format}
\bibliography{jserd}

\end{document}